# SQUARE WAVE EXCITABILITY IN QUANTUM DOT LASERS UNDER OPTICAL INJECTION

M. Dillane[1,2], B. Tykalewicz[2,3], D. Goulding[2,3,4], B. Garbin[5,6], S. Barland[6], and B. Kelleher[1,2]

[1]Department of Physics, University College Cork, Ireland
[2]Tyndall National Institute, University College Cork, Lee Maltings, Dyke Parade, Cork, Ireland
[3]Centre for Advanced Photonics and Process Analysis, Cork Institute of Technology, Cork, Ireland
[4]Department of Mathematics, Cork Institute of Technology, Cork, Ireland
[5]The Dodd-Walls Centre for Photonic and Quantum Technologies, Department of Physics, The University of Auckland, Auckland 1142, New Zealand
[6]Université Côte d'Azur, INPHYNI, Sophia-Antipolis, France

December 12, 2018

ABSTRACT

Quantum dot lasers display many unique dynamic phenomena when optically injected. Bistability has been predicted in a region of high injection strength. We show experimentally, rather than a phase-locked bistability, a square wave phenomenon is observed in this region. The squares can manifest as a periodic train but also as noise driven Type II excitable events. We interpret the appearance of the square waves as a thermally induced breaking of the bistability. Indeed we find experimentally that over the duration of a square the relative detuning between the master and the slave evolves deterministically. A relatively simple, physically motivated, rate equation model is presented and displays excellent agreement with the experiment.

## 1 Introduction

Excitability has been shown to be a pervasive phenomenon in systems of semiconductor optical devices. It has been observed in semiconductor ring lasers [1], semiconductor optical amplifiers [2], devices with saturable absorbers [3] and lasers under feedback [4]. Most importantly for this work, it arises in the optical injection configuration [5]. At low injection strengths Type I excitable pulses are observed and each pulse corresponds to a $2\pi$ phase slip [6]. Single pulse excitability is typical although multipulse generalisations have been observed [7, 8, 9, 10], corresponding to rotations of multiples of $2\pi$. The pulses arise stochastically close to the boundaries of phase locking but importantly from a potential application point of view, they can be deterministically triggered as shown in [11, 12]. The Adler model [13] is the prototype for this behaviour and the underlying bifurcation structure is a saddle node on an invariant cycle [5].

In this work we demonstrate a new excitable phenomenon in the injected quantum dot (QD) laser. The laser under investigation is similar to the one used in [14] and the gain medium consists of InAs QDs grown on a GaAs substrate. The device was operated to emit from the ground state only. For high injection strengths, where the intensity of the injected light is greater than 0.5 times the free-running SL, we find a regime of periodic square waves. Square wave outputs have been observed in several different feedback configurations [15, 16, 17], but they arise from very different underlying physical phenomena to the squares in our work. In our case we find an optothermal dynamic similar to the dynamics observed with semiconductor etalons [18], QD lasers [14, 19, 20], photonic crystal structures [21] and optically injected semiconductor optical amplifiers [2, 22]. Our periodic square wave regime is bounded by regions where individual, square pulses or dropouts are observed. We interpret these individual pulses and dropouts as Type II excitable events akin to those observed in the Van der Pol Fitzhugh Nagumo model [23, 24]. The similar systems mentioned display fast-slow dynamics with two timescales: the timescale of carriers in the nanosecond range and a thermal timescale in the microsecond range. In each case, a bistability is broken by thermally induced dynamics.

For optically injected QD lasers at high injection strengths, a regime of phase-locked bistability has been predicted theoretically [25, 26]. In this letter we show that this bistability is broken by an optothermal coupling, yielding a deterministic square wave cycle.

## 2  Experiment

A single mode InAs based QD laser, pumped at 40 mA (1.25 times threshold), was optically injected using a single mode lensed fiber with a coupling efficiency of approximately 70%, in a unidirectional master slave configuration. The master laser (ML) was a commercial tunable laser source (TLS) with minimum step size of 0.1 pm (0.0178 GHz). Light from the ML was injected into the slave laser (SL) via a circulator and a polarisation controller maximized the coupling. The output was detected with 12 GHz detectors connected to a real-time digital oscilloscope.

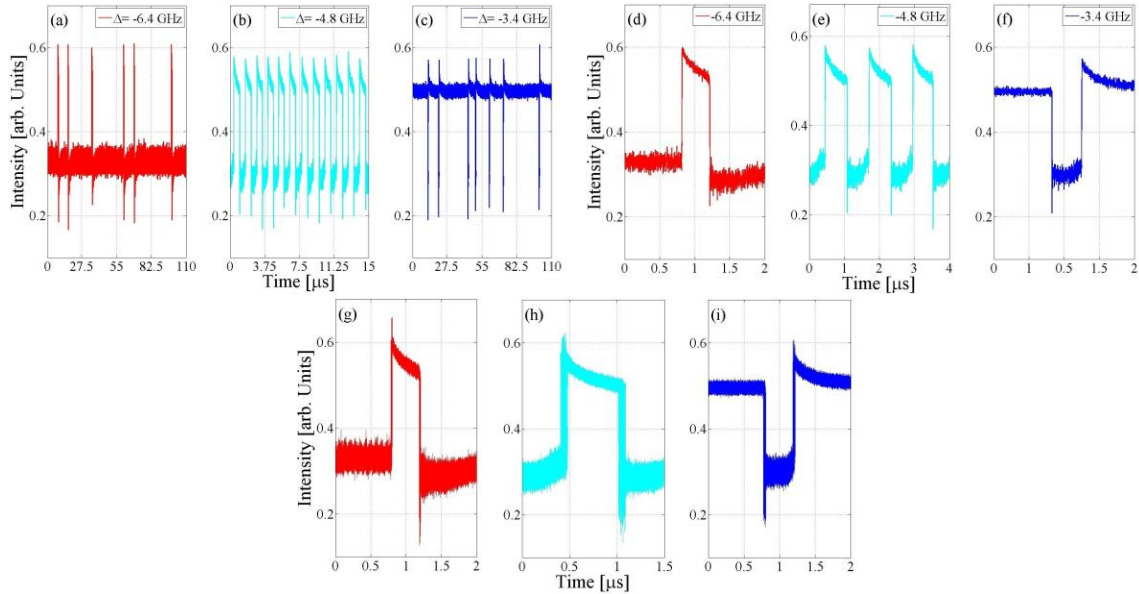

Figure 1: Extracts taken from a 20 ms long timetrace of SL output. (a) randomly separated pulses at -6.4 GHz detuning. (b) periodic squares at -4.8 GHz detuning and (c) square dropouts appearing randomly at -3.4 GHz detuning. The middle row zooms in on a square event from the top row. The bottom row shows a persistence plot where 100 consecutive pulses are overlaid upon each other. Pulse widths are consistent.

For this work, the area of interest is near the unlocking boundary at negative detuning (where the detuning is defined as the linear frequency of the ML minus that of the SL). For low and moderate injection strengths (where the intensity of the injected light is less than 0.5 times the free-running SL) the observations mirrored those reported in previous work [6, 9, 10, 26]. For high injection strengths and for a significant range of negative detuning, a square wave phenomenon was observed.

Beginning where the frequency of the ML was much less than that of the SL, the output of the SL was a limit cycle, where the amplitude decreased as the frequency of the ML was increased until reaching a locking region. Here the output became constant (modulo noise). Increasing the ML frequency to a detuning of approximately -6.6 GHz the first square pulses appear.

Initially, there is a long time between successive pulses. To obtain a sufficient number of pulses for a statistical analysis the detuning had to be further increased to approximately -6.4 GHz. Figure 1(a) shows the randomly spaced square pulses observed at this detuning. Figure 2 shows the distributions of interpulse/interdropout times for several detuning values. The Kramers like exponentially decaying distribution in Figure 2(a) confirms the random, noise induced nature of this train. In this case, the escape time was approximately 6.9 $\mu$s. While the time between the pulses was random, the pulse-width itself was extremely regular, approximately 0.4 $\mu$s. Comparing a persistence plot of 100 successive pulses (Figure 1(g)) to a single pulse (Figure 1(d)) it's clear that each pulse is virtually identical. As the frequency of the ML was brought closer to that of the SL the pulse train became periodic, eventually manifesting with a 50-50 duty cycle at approximately -4.8 GHz detuning (Figure 1(b)). In Figure 2(b) the narrow distribution of high and low intensity lifetimes confirms the periodicity of the train and Figure 1(h) shows the consistent pulse width. Continuing to increase the ML frequency, rather than a pulse train, the train resembled more a periodic train of square dropouts. (That is, the lifetime in the upper level became longer than that in the lower level.) Eventually the dropouts became randomly separated, with an example shown in Figure 1(c). As shown in Figure 2(c), an exponential distribution with an escape time of approximately 6.6 $\mu$s was obtained at a detuning of -3.4 GHz. Similarly to the pulses, the width of each dropout was consistent as shown in the persistence plot in Figure1(i). Finally, further increasing the detuning led to a constant phase-locked solution (modulo noise) at approximately -3.1 GHz. The dynamics resemble those of a Van der Pol/Fitzhugh-Nagumo oscillator and we interpret the pulses and dropouts in the random trains as Type II excitable events. Each escape is noise-induced and is followed by a large deterministic trajectory back to the steady state. The transition between the upper and lower levels of each square is approximately 1 ns and the lifetime in each plateau is on the order of a $\mu$s. We associate this slow-fast, two time scale phenomenon with an optothermal coupling: the fast time corresponds to carrier dynamics and the slow time with thermal effects.

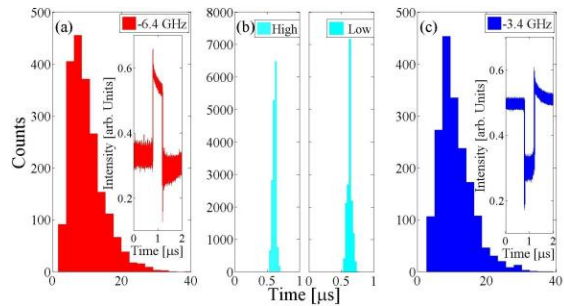

Figure 2: Histograms of the interpulse/dropout times taken over 20 ms. The detuning values correspond to Figure 1. (a) shows the interpulse time distribution of 2,016 pulses. The exponential shape indicates that these pulses are randomly separated and induced by noise. (b) shows the distributions found in the 50-50 duty cycle case in the high (left) and low (right) intensity plateaux. (c) shows the distribution for 1,818 dropouts. Again, the exponential shape is evidence that the drop-outs are noise induced random events. The insets are the persistence plots.

As mentioned above, in [25, 26] a phase locked bistability was predicted theoretically for the system at high injection strength in the region where the square waves are found. We interpret the squares as arising from a breaking of this bistability due to the optothermal coupling. The top and bottom of each square then correspond physically to the two phase-locked solutions. The different intensities and carrier densities of each state lead to a different temperature in the device. Non-radiative recombinations are one of the leading causes of such temperature changes [18]. Changes in temperature lead to changes in refractive index and thus to changes in the effective detuning. We note that in any one square oscillation, the overall temperature change is small (much less than a degree Kelvin). Ultimately this means that depending on the details of the solution, the effective detuning between the master and slave lasers can be different for the same experimental control parameters (the wavelength and the power of the ML). There is a large difference between the intensities of the two states resulting in an appreciable difference in the effective detuning. After a state switch, the associated thermal change moves the system deterministically back in the direction of the original state leading to another state switch after which the process begins again. The hysteresis cycle associated with the bistability is thereby transformed into a deterministic cycle.

To measure the thermally induced change in the detuning, a sub-threshold longitudinal mode was investigated. When free running at 1.25 times threshold, the nearest longitudinal mode of lower frequency happened to be suppressed by only 29 dBm for this device, while all other modes were suppressed by at least 41 dBm. A second TLS slightly negatively detuned from this mode was mixed with the output from the SL. Only during the lower part of the cycle was there sufficient intensity in the side mode to measure the beat signal. We found the evolution of the frequency of this beating cycle by calculating the Fast Fourier Transform of many ten nanosecond intervals. Figure 3(b) shows a slow deterministic change in the frequency of the SL of approximately 2 GHz over the duration of the lower intensity level. This confirms that the effective detuning is indeed changing during each square wave. We show now that the optothermal coupling can be easily incorporated into a simple model for the system and the results are in excellent agreement with the experiment.

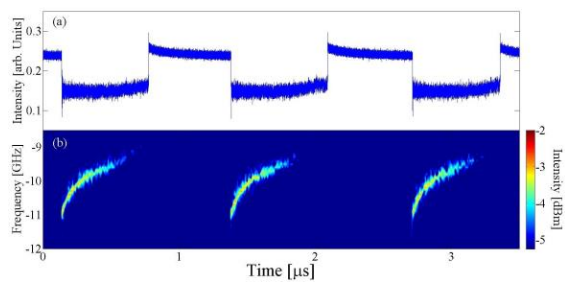

Figure 3: A periodic square wave is shown in (a). A second TLS was mixed with the output of the SL and slightly negatively detuned from a side mode. When the side mode had sufficient intensity (in the lower section of the square) a beat signal was seen. (b) shows the evolution of this beat frequency over time. During the lower section of the square, the slave laser undergoes a deterministic 2 GHz drift due to the optothermal effect.

## 3 Theory

Rate equation models of the optically injected system have proven to be excellent predictors of the experimentally observed dynamics for semiconductor lasers [27]. While specifically tailored QD models do exist, they are necessarily more complicated than their conventional counterparts [25, 28]. It has been shown that qualitatively, one can recover many of the phenomena observed with optically injected QD lasers using the conventional semiconductor rate equations but with very high damping of the relaxation oscillations (ROs) imposed, as found experimentally for QD lasers [29]. In fact, one can even recover virtually all of the phenomena by moving to the Class A, infinite RO damping limit, where the dependence on the carriers is adiabatically eliminated and the behaviour is described using only the electric field [25, 26, 29]. Most importantly here, such lasers display a phase-locked bistability precisely of the form predicted for optically injected QD lasers [25, 26, 29, 30]. Thus, in order to highlight the physics of the phenomenon, we work with this model while recognising that QD lasers are not genuine Class A lasers. Nonetheless, we manage to capture the basic physics and find excellent qualitative agreement.

Next, we modify the model so as to include thermal effects. Carrier heating effects have previously been incorporated in a very detailed microscopic model for QD lasers in [31] and their importance in large signal modulation highlighted. In our case, we want to focus on the underlying physics of the system and so we modify our model to include thermal effects by adding one equation that couples the effective detuning to the intensity. Despite this simplicity, we find that the model is sufficient to understand the essentials of the mechanism. Our model is then

$$\dot{R} = \frac{P - R^2}{1 + 2R^2} R + K \cos\phi \tag{1}$$

$$\dot{\phi} = \alpha \frac{P - R^2}{1 + 2R^2} - (\Delta - \omega) - \frac{K}{R} \sin\phi + \eta \xi(t) \tag{2}$$

$$\dot{\omega} = -\gamma \left(\omega - cR^2\right), \tag{3}$$

where $R$ is the slave field amplitude, $K$ measures the injection strength, $\varphi$ is the phase of the slave minus that of the master, $\Delta$ is the nominal angular detuning, $\alpha$ is the phase amplitude coupling factor and $P$ is the pumping current above threshold. We include stochastic effects via $\eta \xi(t)$, where $\xi(t)$ is a Wiener process and $\eta$ is the magnitude. $\omega$ is the thermal change in the detuning, $\gamma$ gives the slow characteristic time scale for the thermal effects, and $c$ is the coupling between the thermal dynamics and the intensity. Time is expressed in units of the photon lifetime.

The bare ($c = 0$) system, shown in [29], has a region of phase-locked bistability and an associated hysteresis loop (with $K$ in the interval [0.5,0.8], for $\alpha$ = 2 and $P$ = 0.5 for example) due to the interaction of a saddle-node bifurcation and a Hopf bifurcation at negative detuning. No square wave regime is obtained. Consider now the coupled system with finite $c$. Some things remain virtually unchanged. In the limit of low injection strength the Adler equation can be recovered for the phase equation. Thus, Type I excitability is still obtained as experimentally observed. When we move to higher injection strengths and to the previous region of bistability however, things have changed dramatically.

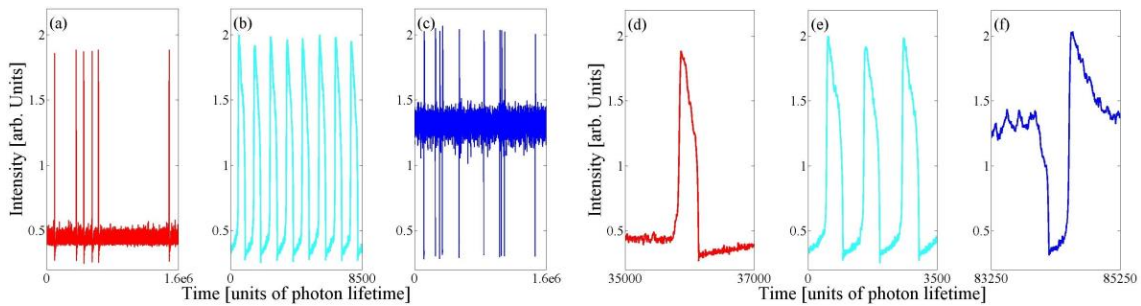

Figure 4: Numerical pulse/dropout trains. (a) shows the intensity at a detuning of -0.955; (b) shows the intensity at a detuning of -0.9225; (c) shows the intensity at a detuning of -0.89. K is 0.7056. The bottom row shows zooms of the top row.

In the absence of noise, a prominent region of periodic square wave pulses is obtained. Close to the onset of this regime, the system is excitable which we confirmed by applying sufficiently large discrete perturbations. To better mimic the experiment we include noise and analyse the evolution of the intensity. For the rest of this Letter we fix $\alpha$ = 2, $P$ = 0.5, $c$ = 0.1, $\gamma$ = 0.001 and $\eta$ = 0.1. Excitable square wave trains are obtained as shown in Figure 4(a) and Figure 4(c).

Figure 4(b) shows an example of the periodic square regime (with noise). The associated interpulse/interdropout histograms are shown in Figure 5 and match the experimental case extremely well.

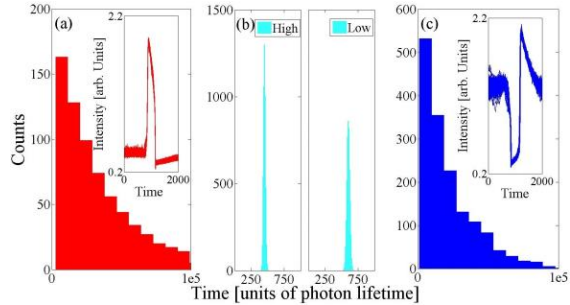

Figure 5: The distribution of the interpulse times is seen in (a) and the inset is a persistence plot of 100 consecutive pulses plotted on top of each other illustrating the regular pulse width. (b) shows the distributions of the high and low intensity states for a periodic train. (c) shows the distribution for the square dropouts with a persistence plot inset. The control parameters match those in Figure 4.

An obvious question is why such behaviour is not observed with conventional semiconductor lasers. Such a phaselocked bistability only arises in experimentally accessible regions for highly damped lasers [29], so the absence of the phenomenon in conventional weakly damped, semiconductor lasers, is natural.

To conclude, we have shown the first demonstration of Type II excitability in the optically injected laser system. At high injection strengths, optothermal effects break a phase-locked bistability and endow the system with a periodic square wave regime and associated regions of Type II excitability. The relative detuning undergoes a deterministic sweep over the duration of a square. Interestingly, this is the first laser system displaying both Type I and Type II excitability. This suggests that QD lasers could be of immense interest in neuromorphic systems where artificial neuronal systems are desired [32, 33, 34], providing prototypical excitable behaviour for both types. Further, we would expect the phenomenon to be robust to changes in device type with the main requirement being high RO damping.